\documentclass[12pt]{article}

\def\be{\begin{equation}}
\def\ee{\end{equation}}
\def\ba{\begin{eqnarray}}
\def\ea{\end{eqnarray}}
\def\half{{1 \over 2}}
\def\tphi{\tilde{\phi}}
\def\e{\mbox{e}}
\def\irho{b}
\input epsf
\begin{document}
\title{Decoherence, instantons, and cosmological horizons}
\author{S. Khlebnikov \vspace{0.1in} \\
{\it \normalsize Department of Physics, Purdue University, West Lafayette, 
IN 47907, USA}}
\date{November 2001}
\maketitle
\begin{abstract}
We consider the possibility that tunneling in a degenerate double-well 
potential in de Sitter spacetime leads to coherent oscillations of 
quantum probability to find the system in a given well. We concentrate 
on the case when the mass scale of the potential is much larger than 
the Hubble parameter and present a procedure for analytical continuation 
of the ``time''-dependent instantons, which allows us to study 
the subsequent real-time evolution. We find that the presence of 
the de Sitter horizon makes tunneling completely incoherent and calculate 
the decoherence time. We discuss the difference between this case and 
the case of a $\theta$-vacuum, when tunneling in de Sitter spacetime 
preserves quantum coherence. \\
\hspace*{3.9in} hep-ph/0111194
\end{abstract}
\section{Introduction}
Tunneling in de Sitter spacetime had been a subject of many papers
\cite{HM}--\cite{GT}.
Our purpose in revisiting it here is to explore its possible
connection with decoherence of quantum fields during cosmological
evolution. Decoherence refers to the onset of classical behavior
in a quantum system or, more formally, to vanishing of the off-diagonal
elements of the reduced density matrix. It is well-known that the flow of
quantum modes of a scalar field through a de Sitter horizon can result 
in an effectively classical behavior of the field's large-scale, 
coarse-grained component, a phenomenon concisely described
in Starobinsky's stochastic approach \cite{Starobinsky}.

In laboratory systems considered as candidates for macroscopic quantum
coherence (MQC) experiments, decoherence is attributed to interactions with
the environment. In this case, it is known that a sufficiently strong such 
interaction can entirely wash out any coherent quantum oscillations associated 
with tunneling \cite{CL}. It is by no means obvious that the special dynamics
of super-Hubble modes in de Sitter space can be interpreted as decoherence 
in the usual MQC sense \cite{Habib}. Nevertheless, one may reasonably ask what 
effect the de Sitter horizon has on various 
specific manifestations of quantum coherence.

One manifestation of quantum coherence, important in particle physics, is
the $\theta$-vacuum structure \cite{theta} in gauge theories, such as QCD.
$\theta$-dependence 
of the partition sum is a result of interference between topologically
distinct paths in the configuration space and has no counterpart 
in classical theory. In theories that are conformally
invariant at the classical level, such as massless gauge theories in
four dimensions, it is easy to construct the corresponding instanton
solutions for the de Sitter case and find a nonvanishing, although---as
explained in the conclusion---probably reduced,
$\theta$ dependence. (This implies in particular that the de Sitter 
expansion alone cannot wash out an unwanted $\theta$ angle.)

Here we want to consider another class of problems where destruction
(or persistence) of quantum coherence is important. These arise in
cosmological scenarios (such as eternal inflation 
\cite{etern}) in which there is an infinite supply of universes, each with 
its own values for at least some of the fundamental constants. 
The question is if there is
a mechanism that causes our universe to be in a state with definite
values for these constants, rather than in a quantum superposition of 
such states. If one accepts that values of the fundamental constants 
are set by expectation values of some scalar fields, e.g. superstring 
moduli, the question can be reformulated in terms of these expectation
values. 

In a scalar theory with multiple potential minima, the tunneling rate between 
the minima will be nonzero in a de Sitter spacetime even if it is zero in the 
flat space. 
An example is provided by a scalar potential with two {\em degenerate}
minima (although many of our conclusions will apply to the nondegenerate 
case as well). Because the Euclidean counterpart
of a de Sitter spacetime is compact (a sphere) \cite{GH}, such a theory always
has finite-action Euclidean solutions---instantons. One such solution is
the Hawking-Moss (HM) instanton \cite{HM}, for which the field is at
a maximum of the potential.

The HM instanton can be used to calculate the rate of transitions between 
different potential minima 
in the case when the mass of the scalar is smaller than some critical value
of order of the Hubble parameter $H$ \cite{HM,JS}. This is precisely the case 
when the amplification of super-Hubble modes leads
to stochastic dynamics of the coarse-grained field \cite{Starobinsky}. The
connection between the two approaches has been discussed in the literature
\cite{Starobinsky,GL}. We have little doubt that in this case the stochastic 
dynamics will cause a rapid decoherence between the minima.

In the present paper, we concentrate on the opposite case, when the field has
a degenerate potential but the mass scale of the potential is much larger than
$H$. Note that for a degenerate (or nearly degenerate) potential the large 
mass does not imply that we are headed towards the 
limit \cite{Coleman&DeLuccia} of small, nearly flat-space instantons. Indeed, 
we have already noted that for an exactly degenerate potential there are no
instantons in the flat space.

In perturbation theory, there is no amplification of super-Hubble modes of
a heavy field, and the usual form of stochastic description does not apply.
One may wonder if in this case tunneling leads to the typical MQC phenomenon of
quantum oscillations of the probability to observe the system near a given minimum.
Our main result is that it does not: the presence of the de Sitter horizon
makes tunneling completely incoherent even for fields that are much heavier than 
$H$.

The plan of the paper is as follows. In Sect. 2, mainly to introduce various 
coordinate systems useful for studying instantons in de Sitter space, we consider
a case when tunneling does exist in the flat spacetime---the $\lambda\phi^4$ 
theory with the ``wrong'' sign of $\lambda$. This theory is classically 
conformally invariant, and the instanton is easy to find: it is a conformally
transformed Fubini-Lipatov \cite{FL} instanton. This solution has the conventional
large- and small-size limits.

In Sect. 3 we introduce the $\phi^4$ theory with a degenerate double-well 
potential and discuss the crossover in the properties of the lowest-action
instanton solution at $\mu=\mu_{\rm cr} \sim H$. As explained above, 
the small-instanton limit does not exist in this case, so at $\mu \gg H$, the
lowest-action solution is a different brand of ``time''-dependent instanton:
in the ``static'' coordinates, its ``time'' duration is order $\mu^{-1}$, while 
the spatial extent is of order
$H^{-1}$. In Sects. 2 and 3 we obtain several explicit expressions for the
turning points of the ``time''-dependent solutions. The existence of turning 
points is indeed their key feature, which makes these solutions analogous 
to {\em periodic instantons} \cite{periodic} in massive gauge theories in flat
space (while the HM instanton, from this point of view, is analogous to the 
sphaleron \cite{sphal}).

Sects. 4 and 5 contain the main results of this paper. In Sect. 4 we describe
an analytical continuation of the ``time''-dependent 
solutions that allows us to consider the subsequent real-time evolution.
We find that by the time the field that started at a turning point reaches
a section of constant conformal time $\eta = \eta_0$, it turns into
a super-Hubble droplet of the new phase 
(e.g. $\phi = \mu$ nucleated inside the phase with $\phi = -\mu$).
We argue from causality and confirm by numerical integration
that in de Sitter spacetime such a droplet will expand forever, even if the
corresponding flat-space theory does not have any expanding droplets 
(as is the 
case for a strictly degenerate potential). In Sect. 5 we show 
that, as a result,
the tunneling between $\phi = \mu$ and $\phi = -\mu$ is completely incoherent
and calculate the corresponding decoherence time. Sect. 6 is a conclusion.

Throughout the paper, we neglect backreaction of instantons on the metric. This
is a good approximation when the variation of the potential during an
instanton transition is much smaller than $H^2 M_{\rm Pl}^2$. For example, for
the double potential, the variation is at most of order $\mu^4/\lambda$, where
$\lambda \ll 1$ is the coupling constant. To consider $\mu \gg H$, we then
have to require $H^2 /\lambda \ll M_{\rm Pl}^2$, a relatively mild restriction.

\newpage
\section{Conformally invariant model}
\subsection{Instanton solution}
Our first example is the massless $\lambda\phi^4$ theory with a 
``wrong'' (negative) sign of $\lambda$, conformally coupled to gravity. The
Euclidean action of the field is
\be
S_E = \frac{1}{|\lambda|} \int d^4 x \sqrt{g} \left\{
\half g^{\mu\nu} \partial_\mu \phi \partial_\nu \phi +
{1\over 12} R \phi^2 - {1\over 4} \phi^4 \right\} \; ,
\label{S1}
\ee
where $g_{\mu\nu}$ is the metric, and $R$ is the scalar curvature of a
four-sphere (the Euclidean counterpart of the de Sitter space \cite{GH}). 
The radius of the sphere equals the Hubble radius, which in this section 
we will set to one by a choice of the length unit.
Then, $R=12$.

There are several convenient choices of metric on a unit four-sphere.
One is the conformal form
\be
g_{\mu\nu}(x) = \frac{4\delta_{\mu\nu}}{(1 + x^2)^2} \; ,
\label{met1}
\ee
where $x$ is a flat-space four-vector, and $x^2$ is its length squared.
Thus, the sphere is related to the flat space by a conformal transformation
\be
g_{\mu\nu}(x) \to {\tilde g}_{\mu\nu}(x) = \gamma^2(x) g_{\mu\nu}(x)
\label{trang}
\ee
with
\be
\gamma(x) = (1 + x^2) / 2 \; .
\label{gam}
\ee
The action is invariant under transformations (\ref{trang}) provided we 
also transform the field:
\be
\phi(x) \to {\tilde \phi}(x) = \gamma^{-1}(x) \phi(x) \; ,
\label{tranp}
\ee
and use the scalar curvature corresponding to the new metric: $R\to {\tilde R}$.

The theory in the flat space has a well-known solution, the Fubini-Lipatov
instanton \cite{FL}, which describes decay of the false vacuum $\tphi = 0$:
\be
{\tilde \phi}(x) = \frac{2\sqrt{2} \irho}{(x-x_0)^2 + \irho^2} \; ,
\label{sol_flat}
\ee
where $\irho$ and $x_0$ are arbitrary (real) parameters. Inverting 
the transformation (\ref{tranp}), with $\gamma$ given by (\ref{gam}), 
we obtain the corresponding solution on the sphere
\be
\phi(x) = \frac{\sqrt{2} \irho  (1 + x^2)}{(x-x_0)^2 + \irho^2} \; .
\label{sol}
\ee

This solution was easy to find in the conformal coordinates (\ref{met1}), 
but to study some of its properties it is convenient to consider other 
coordinate systems, which we will also use in the following sections. One of
these new choices is the system of four spherical angles 
$(\zeta,\psi,\theta,\phi)$, which is most easily introduced if we first go
from the coordinates $x$ to the corresponding spherical coordinates, by
writing the line element as
\be
ds^2 = g_{\mu\nu} dx^\mu dx^\nu = \frac{4}{(1+{\cal R}^2)^2} 
\left( d{\cal R}^2 + {\cal R}^2 d\Omega_3^2 \right) \; ,
\label{met2}
\ee
where ${\cal R}$ is the length of $x$, and $d\Omega_3^2$ is the metric on 
a unit three-sphere:
\be
d\Omega_3^2 = d\psi^2 + \sin^2\psi~d\theta^2 + \sin^2\psi \sin^2\theta~d\phi^2 \; .
\ee
We then define the fourth angle $\zeta$ via
\be
{\cal R} = \tan (\zeta/2) \; .
\label{R}
\ee
The line element becomes $ds^2 = d\zeta^2 + \sin^2\zeta d\Omega_3^2$.

The form of solution (\ref{sol}) in the new coordinates
is somewhat complicated, except for the case $x_0 = 0$, on which we now 
concentrate. (This implies no loss of generality for our present purposes, 
as we are free to
center the stereographic projection (\ref{met1}) on the instanton.) Then,
\be
\phi(x) = 
\frac{\sqrt{2} \irho}{\sin^2 (\zeta/2) + \irho^2 \cos^2 (\zeta/2) } \; .
\label{sol2}
\ee
We can see how the solution
changes as we change $\irho$ from $\irho \ll 1$ to $\irho \gg 1$.
At small $\irho$, the solution is concentrated in a small region near $\zeta = 0$ 
(the north pole of the four-sphere), where
\be
\phi \approx \frac{4\sqrt{2} \irho} {\zeta^2 + 4\irho^2} \; .
\ee
This is equivalent to the flat-space instanton (\ref{sol_flat}) with a rescaled
size: $\irho \to 2 \irho$. As $\irho$ increases, the solution occupies a larger
region, until at $\irho = 1$ it becomes uniform over the sphere. This is, of course,
the Hawking-Moss instanton of the model:
\be
\phi_0 = \sqrt{2} \; .
\ee
Any solution with $\irho > 1$ can be obtained from
a certain solution with $\irho < 1$ by using the transformation $\irho\to 1/\irho$, 
$\zeta \to \pi - \zeta$. Thus, at large $\irho$ the solution is concentrated in
a small region near the south pole and is approximated by the flat-space instanton
with $\irho \to 2/ \irho$. Therefore, both large and small $\irho$ limits correspond
to instantons of sizes much smaller than the curvature radius. Such instantons 
are practically unaffected by the
curvature (the Coleman-De Luccia limit \cite{Coleman&DeLuccia}). 
The point $\irho = 1$ corresponds to the largest size and
strong curvature effects.

\subsection{The static coordinates}
\begin{figure}
\leavevmode\epsfysize=3.0in \epsfbox{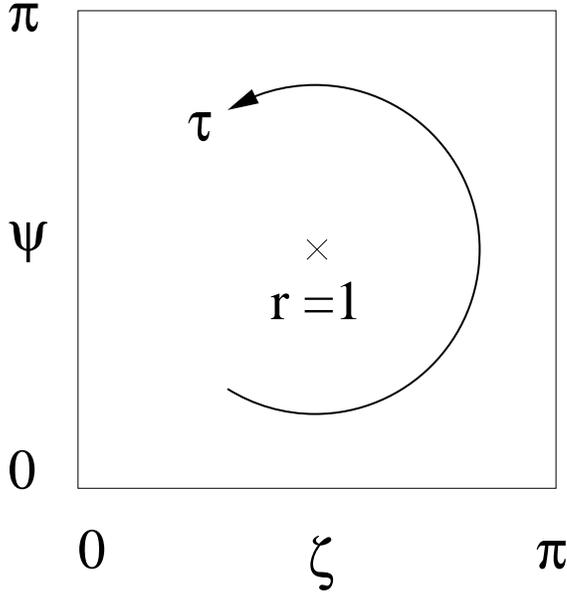}
\vspace*{0.2in}
\caption{Coordinates $r$ and $\tau$ defined by (\ref{r}), (\ref{tau}). The
boundary of the square corresponds to $r=0$, and the center to $r=1$.
}
\label{fig:map}
\end{figure}

Of particular interest is the form of the solution in the so-called ``static''
coordinates, which are obtained by leaving
$\theta$ and $\phi$ of the previous set intact but replacing $\zeta$ and
$\psi$ with
\ba
r & = & \sin\psi \sin\zeta \; , \label{r} \\
\tau & = & \arctan(\cos\psi \tan\zeta) \label{tau} \; .
\ea
The line element now takes the form
\be
ds^2 = (1-r^2) d\tau^2 + \frac{dr^2}{1-r^2} + r^2 d\Omega_2^2 \; ,
\label{met3}
\ee
where $ d\Omega_2^2 = d\theta^2 + \sin^2\theta~d\phi^2$ is the metric of
a unit two-sphere. Each of the angles $\zeta$ and $\psi$ runs from 0 to $\pi$.
For the mapping (\ref{r}), (\ref{tau}) to be continuous, $\tau$ should be 
periodic with period $2\pi$; $r$ runs from 0 to 1. To help visualize this 
mapping we show it pictorially in Fig. \ref{fig:map}.

Metric (\ref{met3}) is ``static'', i.e. independent of $\tau$, so it can be
easily continued to real time. The simplest way to do that
is to make $\tau$ purely imaginary, while leaving $r$ untouched. Although
this choice of coordinates is not convenient for studying the real-time
evolution, the possibility to make it shows that turning points with respect to
$\tau$, i.e. configurations satisfying
\be
\partial_\tau \phi(r,\tau = \tau_0) = 0 
\ee
for some $\tau_0$ and all $r$, are of special importance, since they can be used 
as initial conditions for the real-time evolution (cf. ref. \cite{Deruelle}).

Solution (\ref{sol}) in the static coordinates takes the form
\be
\phi(x) = 
\frac{2\sqrt{2} \irho}{(\irho^2 + 1) + (\irho^2 -1) \sqrt{1-r^2} \cos\tau} \; .
\label{inst}
\ee
We see that for any $\irho$ different from 1, the instanton oscillates
in the Euclidean time $\tau$, and there are two turning points, at $\tau = 0$ 
and $\tau = \pi$.

When $\irho$ is close to 1, we can expand (\ref{inst}) in small 
$ \irho^2 - 1 \equiv 2\epsilon$: 
\be
\phi(x) = \phi_0 \left( 1 - \epsilon  \sqrt{1-r^2} \cos\tau \right) 
+ O(\epsilon^2) \; .
\ee
So, in this limit, the oscillations are small harmonic oscillations about 
$\phi =\phi_0$.

Because the model (\ref{S1}) is classically conformally invariant, the Euclidean
action will have the same value for any value of $\irho$. Which 
values dominate the false vacuum decay will be decided by quantum
correction to the amplitude. (Because the $\lambda\phi^4$ theory with a negative
$\lambda$ is asymptotically free, we expect that the quantum corrections will
favor the largest instanton sizes, i.e. $\irho \approx 1$.) Certain values of
$\irho$ will 
be favored classically if we deform the model away from the exact conformal 
invariance, by adding a mass term $m^2_{\rm eff} \phi^2 /2$ inside the brackets 
in (\ref{S1}).  
Here
\be
m^2_{\rm eff} = m^2 + (\xi - 1/6) R \; ,
\ee
and $m^2$ and $\xi$ are parameters. 
If $|m^2_{\rm eff}| \ll H^2$, the change in the action can be computed using the
instanton solutions of the conformally invariant theory, cf. ref. \cite{tHooft}.
One can then show that $m^2_{\rm eff} > 0$ favors
instantons with $\irho \to 0$ and $\irho \to \infty$, while $m^2_{\rm eff} < 0$
those with $\irho\approx 1$. This is a variant of the well-known crossover \cite{HM,JS}
in the properties of the dominant classical solutions, and we do not go into 
details of the calculation here.

\section{Double-well potential}
We now turn to models in which, unlike the model of the previous section,
there are two degenerate or nearly degenerate minima. In the next two sections,
we will try 
to answer the main question of this paper, namely, whether in such cases
tunneling in de Sitter space can lead to coherent oscillations of the probability 
to find the system near a given potential minimum.

The simplest model of this kind is a scalar $\phi^4$ theory with an exactly 
degenerate double-well potential. The Euclidean action of the field is
\be
S_E = {1\over \lambda} \int d^4 x \sqrt{g} \left\{
\half g^{\mu\nu} \partial_\mu \phi \partial_\nu \phi +
\half \xi R \phi^2 - \half \mu_0^2 \phi^2 + {1\over 4} \phi^4 \right\} \; ,
\label{S2}
\ee
where $\lambda$ is now positive, and the parameter $\xi$ can take any value.
As before, we measure lengths in units of the Hubble radius, so $R=12$.
The $\phi^2$ terms in (\ref{S2}) can be assembled into a single $\phi^2$ term,
with the mass parameter
\be
\mu^2 = \mu_0^2 - 12 \xi \; .
\ee
We will assume that $\mu^2$ is positive.

In this section, we will use metric in the form (\ref{met2}), 
with an additional definition
\be
y = \ln {\cal R} \; .
\label{y}
\ee
Thus, the line element is
\be
ds^2 = \frac{1}{\cosh^2 y} \left( dy^2 + d\Omega_3^2 \right) \; .
\label{met4}
\ee
(which is conformally equivalent to the metric of a cylinder). 

We now look for $O(4)$ invariant instantons of this theory, i.e. solutions
for which $\phi$ depends only on $y$. In view of the relations
\be
\tanh y = -\cos\zeta = -\sqrt{1 - r^2} \cos\tau \; ,
\label{cos_zeta}
\ee
following from (\ref{R}),
(\ref{y}), (\ref{r}), and (\ref{tau}), we expect any
such solution to have at least two turning points with respect
to the Euclidean time $\tau$, at $\tau=0$ and $\tau=\pi$. 
Equation of motion for such solutions is 
obtained from (\ref{S2}) and reads
\be
-\partial_y^2 \phi + 2\tanh{y}~\partial_y\phi + {1\over \cosh^2{y}} \left(
- \mu^2 \phi + \phi^3 \right) = 0 \; .
\label{eqm2}
\ee 
It is useful to consider also the field $\tphi = \phi / \cosh{y}$, for which
the equation has no ``friction'' term:
\be
-\partial_y^2 \tilde{\phi} + \tilde{\phi} 
- \frac{\mu^2 + 2}{\cosh^2{y}}~\tilde{\phi} + \tilde{\phi}^3 = 0 \; .
\label{eqm1}
\ee
For solutions describing tunneling, we require that $\phi$ goes to constant
values at $y\to\pm\infty$, and these constant values are different.

At large $\mu$, eq. (\ref{eqm2}) has a solution for which $\phi$ changes
from $\phi \approx -\mu$ to $\phi \approx \mu$ in a narrow (of order $1/\mu$) region
near $y = 0$. This solution is well approximated by the kink of double-well
quantum mechanics:
\be
\phi \approx  \mu \tanh (\mu y /\sqrt{2}) \; .
\label{kink}
\ee
Because of the finite volume of the de Sitter four-sphere, this solution
has a finite Euclidean action, of order $\mu^3 / \lambda$. On the other
hand, if we 
formally allow $\mu^2$ to take negative values, the usual mechanical analogy,
in which $\tphi$ is viewed as position of a ``particle'', and $y$ as a
fictitious ``time'', shows that for $\mu^2 \leq -1$, eq. (\ref{eqm1}) cannot
have any bounded solutions except $\tphi=0$. We conclude that a $y$-dependent
solution must appear at some critical value $\mu^2 = \mu_{\rm cr}^2 > -1$.

This critical (bifurcation) point can be found by the standard argument \cite{JS},
which we briefly review here. We assume that at $\mu^2$ just
above $\mu_{\rm cr}^2$ the new solution is confined to a vicinity of $\phi = 0$. 
Then, we can linearize eq. (\ref{eqm2}) into
\be
-\partial_y^2 \phi + 2\tanh{y}~\partial_y\phi 
- {{\mu}^2 \over \cosh^2{y}} \phi = 0 \; .
\label{lin}
\ee
For $\mu^2 = 4$, this equation has a solution:
\be
\phi = A \tanh y \; ,
\label{smamp}
\ee
with an arbitrary amplitude $A$ (which should be small to justify
using (\ref{lin}) in the first place), suggesting that
\be
\mu_{\rm cr}^2 = 4 \; .
\label{mucr}
\ee
A numerical study of the full eq. (\ref{eqm1}) confirms this suggestion:
for $\mu^2 \leq 4$,
we have found no solutions with a nonzero value of ``velocity''
$\partial_y{\tphi}$ at $y=0$, while for $\mu^2 > 4$ such 
solutions exist, with the field $\phi$ approaching (\ref{kink}) at large 
$\mu$. We have also verified numerically that the $y$-dependent solution has
a lower action than the Hawking-Moss instanton (which in this model
is $\phi_0 = 0$).

In terms of the static coordinates (\ref{r}), (\ref{tau}), the small-amplitude
solution (\ref{smamp}) has the form
\be
\phi = - A \sqrt{1-r^2} \cos\tau \; ,
\label{smamp2}
\ee
corresponding again to small harmonic oscillations near the Hawking-Moss 
instanton.

Condition (\ref{mucr}) is a special case of the condition
\be
-V''_{\rm cr}(\phi_0)  = 4 \; ,
\ee
obtained in refs. \cite{HM,JS} for a wide class of potentials. Here $V''$ is
the second derivative of the potential, computed on the HM instanton $\phi_0$.
The difference with the strongly supercooled cases considered in \cite{HM,JS}
and in the previous section
is that in the present case the limit $\mu \gg 1$ leads not
to a small, nearly flat-space instanton but to a solution like (\ref{kink}), 
for which the ``time'' duration (in $\tau$) is indeed small, of order $\mu^{-1}$,
but the spatial extent (in $r$) is still large, of order of the Hubble radius.

\section{Continuation to real time}

The existence of two turning points, with respect to ``time'' $\tau$, 
for any $O(4)$ symmetric ``time''-dependent instantons suggests that these 
instantons can be 
analytically continued to the Lorentzian signature, and the analytic 
continuation will describe some real-time process. Continuation of time $\tau$
alone, leaving the radial coordinate intact,
is not convenient for studying the real-time evolution, for a reason that will 
be explained shortly.
So, we make instead a transformation that acts on {\em both} 
$r$ and $\tau$, and such that the new time variable is the conventional conformal
time $\eta$. (Using conformal
time for specifying the {\em pre}-tunneling quantum state was advocated
in ref. \cite{RS}.)

The required transformation of coordinates is
\ba
r & = & - \rho / \eta \; , \label{rho} \\
\tau & = & -\frac{i}{2} \ln(\eta^2 - \rho^2) \; . \label{eta} 
\ea
where $\rho$ is a new radial coordinate. The metric takes the form
\be
ds^2 = {1\over \eta^2} \left( -d\eta^2 + d\rho^2 + \rho^2 d\Omega_2^2 \right) \; ,
\label{met_conf}
\ee
which is the usual real-time conformal metric of de Sitter space. On the real-time
segment of the evolution, $\eta$ can take any value from $-\infty$ to $0$. On
the Euclidean segment, $\eta$ is complex (as it should be; note that $\rho$ is
also complex there), but both $\eta$ and $\rho$ become real at the turning points
$\tau = 0$ and $\tau = \pi$.

At the turning points, eq. (\ref{eta}) gives
\be
\eta^2 - \rho^2 = 1 \; ,
\label{hyper}
\ee
which defines a hyperbola in the $(\eta,\rho)$ plane. To each point on this
hyperbola, one can associate via (\ref{rho}) a unique value of the static
coordinate $r$, satisfying $0\leq r < 1$, see Fig. 2. 
This means that periodic instantons, such as (\ref{inst}), (\ref{kink}), 
or (\ref{smamp2}), completely determine the field and its
derivatives on the hyperbola and thus supply initial data for 
the real-time problem.

\begin{figure}
\leavevmode\epsfysize=3.0in \epsfbox{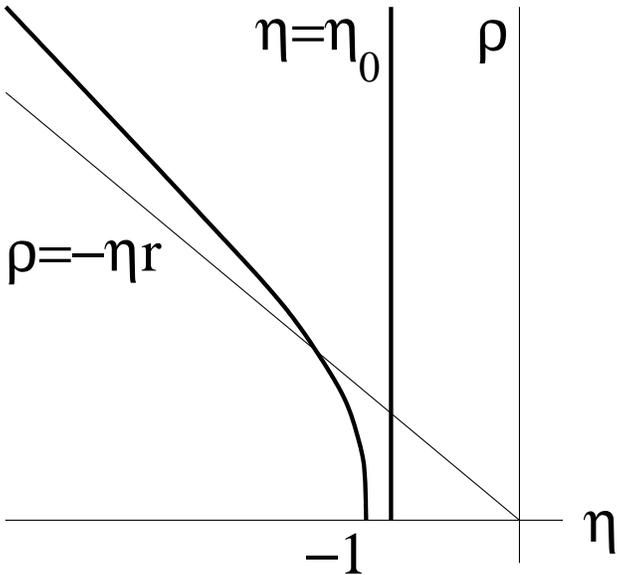}
\vspace*{0.2in}
\caption{Hyperbola (\ref{hyper}), on which the field is at a turning point, and
a section of constant conformal time $\eta$. Evolution of the field between these
two sections is discussed in the text. The line $\rho = -\eta r$
associates a value of the radial coordinate $r$ of the static coordinate system
to a given point on the hyperbola.
}
\label{fig:hyp}
\end{figure}

Note that $-i\tau$, which would be the time variable if we
were to continue $\tau$ alone, 
goes to infinity at the line $\rho = -\eta$. It is for this reason that such
analytical continuation is not convenient for studying the real-time evolution.
It could still be used, though, if we were interested only in the immediate vicinity
of the hyperbola (\ref{hyper}).

It would be very interesting to run the real-time evolution for systems like
those discussed in the previous sections, with the initial data on the hyperbola
determined by the turning points that we have found.
For our present purposes, however, we need only
some general features of the evolution, and these can deduced by relatively simple
arguments. In particular, we will be interested in the general shape of the field
at a constant $\eta$ section, $\eta = \eta_0$ (the vertical line in Fig. 2).

Indeed, the turning points of (\ref{inst}), (\ref{kink}), 
or (\ref{smamp2}) do not look anything like droplets or bubbles of the new phase.
For example, the turning points we have obtained for the double-well potential
do not interpolate between, say, $\phi > 0$ at $r\to 0$ and $\phi < 0$ at
$r\to 1$. Instead, they all tend to $\phi = 0$ at $r\to 1$ (i.e. on the large-$\rho$
part of the hyperbola). We will see, however, that the field acquires the
expected form by the time it reaches $\eta =\eta_0$.

For definiteness, we take $\eta_0\approx -1$, which is near the top of the hyperbola.
Then, for $\rho$ close to zero the field is more or less the same as it was
at the turning point, because the small $\rho$ region has been evolving only
for a short time.
The situation is entirely different for large values of $\rho$. These regions have
evolved for a long time before they arrived at $\eta = \eta_0$, and to find
how the field looks there, we have to consider not only the turning point itself
but also its derivative with respect to $\eta$. (This derivative of course does not
vanish at the turning point---it is the derivative with respect to $\tau$ that does.)

For example, the small-amplitude solution (\ref{smamp2})
in the new coordinates has the form
\be
\phi = - \frac{A}{2\sqrt{\eta^2}} \left (\eta^2 - \rho^2 + 1 \right) \; .
\label{smamp3}
\ee
We need to keep both branches of the square root if want to include both
turning points, because $\eta^2$ changes its phase by $2\pi$ during a half-period
of the periodic instanton. As far as the real-time evolution is concerned, however,
we need to consider only one of the turning points, and only real $\eta$.
Then, it is sufficient to pick only the branch 
corresponding to the turning point in question.
In what follows we use $\sqrt{\eta^2} = \eta < 0$. Then, the solution (\ref{smamp3})
in the vicinity of the hyperbola (\ref{hyper}) is approximately 
$\phi \approx - A / \eta$ at large $|\eta|$, while its derivative is
$\partial_\eta \phi \approx - A$. Note that, unlike the field, the derivative does not
decrease with $|\eta|$ and is of the opposite sign. In other words, in the regions
with large $\rho$ the field initially sits close to the top of the barrier but
is pushed over the barrier, to the ``wrong'' side. The field
in these regions will therefore relax to the ``wrong'' minimum. 

We conclude that the field that
was at a turning point on the hyperbola (\ref{hyper}) becomes a droplet of the
new phase at $\eta =\eta_0$,
i.e. there is a region with $\phi > 0$ surrounded by $\phi < 0$ (or vice versa).
For the case $\mu \gg H$, which is of main interest to us in this paper, we
expect that the field inside and outside the droplet is close to
$\pm \mu$. The size of the droplet in this case is much larger than the Hubble
radius. Indeed, the
large-$\mu$ solution (\ref{kink}) near the part of the hyperbola corresponding
to large $|\eta|$ is approximately of the form (\ref{smamp3}) but with $A$ replaced
by $\mu^2/\sqrt{2}$. To end up on the ``wrong'' side of the potential (in this
case, at $\phi < 0$), the field has to overcome a potential barrier of energy
density proportional to $\mu^2 a^2 \phi^2 \approx \mu^6 / 2\eta^4$, 
where $a$ is the scale factor, while the available
kinetic energy is proportional to $(\partial_\eta \phi)^2 \approx \mu^4/2$.
Thus, the field can get to $\phi < 0$ only in regions that started out 
at $|\eta| \sim \sqrt{\mu}$ or larger, i.e. the size of the droplet (in $\rho$) is 
of order $\sqrt{\mu} / H^{3/2}$. As seen from (\ref{met_conf}), at $\eta = -1$
the scale factor is equal to 1, so the above estimate should be interpreted 
as the physical size at which the droplet nucleates.

Since the argument leading to this picture relied on the existence
of a nonzero derivative with respect to $\eta$ of the analytically continued
instanton, one cannot help wondering if this argument can be adapted to the Hawking-Moss
case, when all the derivatives are zero. In that case, we do not expect any definite
droplet to emerge at $\eta = \eta_0$, because for $\mu < \mu_{\rm cr} \sim H$
(when the HM instanton is relevant) the stochastic dynamics \cite{Starobinsky} will
lead to a random walk of the field at super-Hubble scales. So, instead of talking
about a specific droplet, we should be talking about the statistics of a random
field. If we consider initial conditions for which the field
is equal to the HM instanton on the hyperbola (\ref{hyper}), but allow for a small 
nonvanishing ``velocity'' (a derivative with respect to $\eta$), the above 
argument will go through. Moreover, it may be
possible to obtain the statistics of the random field at $\eta = \eta_0$ from a 
suitable statistics of the small velocity, much like one obtains a
nucleation rate in flat space by summing over all different velocities the
system may have at the top of the barrier. Here, however, we do not pursue this 
idea any further.

In the strongly supercooled case of Sect. 2, the droplet 
will turn into an expanding bubble of the new phase.\footnote{
Adding a sufficiently large constant to the potential in (\ref{S1}) will allow
the field $\phi$ itself to drive a (quasi) de Sitter expansion. In that case, 
the bubble is
precisely of the type contemplated in the new inflationary scenario 
\cite{ninf_Linde,ninf_AS}.}
We now need to develop the corresponding picture for degenerate 
(or nearly degenerate) potentials.
As before, we concentrate on the case $\mu \gg H$, when transitions between
$\phi=\pm \mu$ in a given Hubble volume
are due to tunneling. (In the case $\mu<\mu_{\rm cr} \sim H$, 
they can be regarded as overbarrier transitions, due to the ``Brownian motion'' 
of the field's large-scale component \cite{Starobinsky}.)

In flat spacetime, for a degenerate double-well potential such as (\ref{S2}), 
any droplet of the phase
$\phi = \mu$ nucleated inside the phase $\phi=-\mu$ will collapse
due to the surface tension. We define a {\em nearly} degenerate potential by the 
condition
\be
R_{\rm crit} \ll H^{-1} \; ,
\label{crit}
\ee
where $R_{\rm crit}$ is the size of the critical droplet in flat space;
if a flat-space droplet is larger than the critical size it will expand.
In de Sitter space, a droplet with a size much smaller than the Hubble radius
behaves essentially as its flat-space counterpart. So, in either of the above cases,
such as a droplet will collapse. We have seen, however, that for $\mu \gg H$, 
the size of the droplet is much larger than the Hubble radius. For such a droplet,
the surface is already causally disconnected from the center. By causality,
such a droplet cannot collapse. We therefore expect (and will confirm
by a numerical calculation) that it will develop into an expanding bubble of the 
new phase, much like its counterpart in the supercooled case. (The difference 
remains that the spatial size of the droplet is now strongly dependent on
the Hubble radius.) 

The persistence of super-Hubble droplets in the case of a degenerate potential 
can be directly tested by a
numerical integration of the real-time equation of motion
\be
\phi_{,\eta\eta} + 2{\cal H} \phi_{,\eta} 
-\frac{ (\rho^2 \phi_{,\rho})_{,\rho} }{\rho^2} + a^2 V'(\phi) = 0 \; ,
\label{realtime}
\ee
where $a = - 1/ H\eta$ is the scale factor,
and ${\cal H} = a H$ is the comoving Hubble parameter. For the model
(\ref{S2}) the potential is
\be
V(\phi) = -\half \mu^2 + {1\over 4} \phi^4 \; .
\label{Vphi}
\ee

Fig. 3 presents results of numerical integration of eq. (\ref{realtime}) with
the initial conditions
\ba
\phi(\rho, \eta = \eta_0) & = & 2 \exp(-\rho^2 / \ell^2) - 1 \; , \label{ini1} \\
\phi_{,\eta}(\rho, \eta = \eta_0) & = & 0 \; \label{ini2} .
\label{ini}
\ea
These initial conditions contain a single
parameter ${\ell}$, which regulates the spatial size of the droplet.
Although we have seen that for $\mu \gg H$ droplet sizes
are quite respectably super-Hubble, to prove the point (i.e. the relevance
of the Hubble radius, as implied by the causality argument), we use for Fig. 3 
a mildly super-Hubble value $\ell = 2 H^{-1}$. 
We see that the
{\em comoving} size of the configuration rapidly freezes in, which means 
that in the physical space the size grows at the rate of the expansion. 
In contrast, sub-Hubble droplets (not shown in the plot) collapse.

\begin{figure}
\leavevmode\epsfysize=3.0in \epsfbox{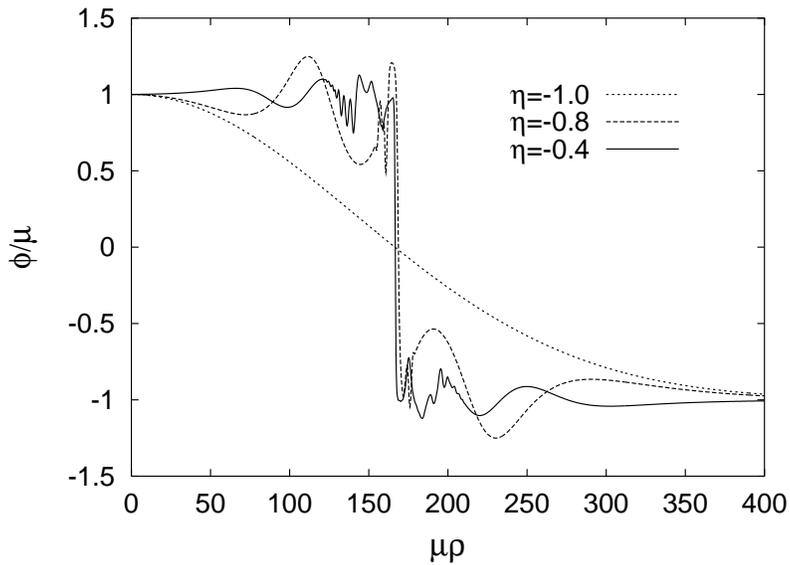}
\vspace*{0.2in}
\caption{Radial profiles of the field $\phi$ at different moments of conformal
time, obtained by numerical integration of eq. (\ref{realtime}) with the potential
(\ref{Vphi}) and initial conditions (\ref{ini1}), (\ref{ini2}), for $\mu = 100$, 
$\ell = 2$, and $\eta_0 = -1$ (all in units of the Hubble rate/radius). 
Because $\rho$ is a comoving coordinate, the position of the Hubble
radius in the plot shifts with time to smaller $\rho$, according to 
$\mu\rho_H = \mu|\eta|$.
}
\label{fig:field}
\end{figure}

The role of the de Sitter horizon is emphasized by a comparison
of these results with the behavior of the same model
in a radiation-dominated (RD) universe. In the latter case, we expect that
even if a droplet starts out as super-Hubble, it will eventually become 
sub-Hubble, as the growing Hubble radius catches up with the droplet's size.
Once the size of the droplet becomes much smaller than the 
value of the Hubble radius at that time, the droplet's evolution is essentially
the same as in flat spacetime. Thus, in an RD universe 
the droplet will eventually collapse regardless of its initial size. 
These expectations are well born out in numerical integrations
of eq. (\ref{realtime}) with $a = H_0\eta +1$ and ${\cal H} = (\eta + 1/H_0)^{-1}$.

The causality argument implies that in a de Sitter spacetime even a droplet 
of false vacuum, in a model with nondegenerate minima, should expand, as long it
has nucleated with a super-Hubble size. This corresponds to the
so-called ``jumps up'' (in potential energy), which are known to occur in de 
Sitter space both for small $\mu$ \cite{jumps_up} and for $\mu>\mu_{\rm cr}$ 
\cite{RS}. Our integrations of (\ref{realtime}), with a nondegenerate
$V$ and $\mu\gg H$, confirm that the real-time picture implied by causality 
is correct also in this case, although now we find that the expansion 
of the droplet, at least initially, proceeds at a slower rate than the 
expansion of the universe.

\section{Calculation of decoherence time}
We now have the tools required to understand the influence of 
a de Sitter horizon on quantum coherence, for fields with $\mu \gg H$.  
If $\phi$ with the potential (\ref{Vphi}) were not a field
but a single quantum-mechanical degree of freedom, tunneling would lead to
quantum oscillations of $P(t) = P_L(t) - P_R(t)$, where $P_{L(R)}$ 
is the probability to observe a negative (positive) $\phi$, i.e. to find the
system in the left (right) well. These oscillations would be of the form
\be
P(t) = \cos\Omega t
\label{osc}
\ee
with $\Omega$ proportional to the tunneling amplitude.

Now that $\phi$ is actually a quantum field in a de Sitter spacetime, the
picture is quite different. As we have seen, each tunneling event nucleates
a super-Hubble droplet of the new phase, and the droplet subsequently expands.
As a result, the state of the field at the center of the droplet changes, say,
from $\phi \approx -\mu$ to $\phi \approx \mu$.
Later, another tunneling event will occur in the same Hubble volume,
with another expanding droplet nucleating, and the field changing back to
$\phi \approx -\mu$. Although within the given Hubble volume the final field
becomes indistinguishable from the original, the super-Hubble
configurations are in most cases quite distinct: if the second droplet has 
nucleated not too soon after the first, there are now two receding bubble walls, 
while originally there were none. If, in such a case, we consider
an overlap of the initial quantum state with the final one, by taking a product 
of individual overlaps for all the field modes, the rapidly growing difference
between the states of super-Hubble modes will lead to a rapid
decrease in the overlap. One consequence of that will be suppression of
quantum oscillations of the probability (decoherence).

To estimate how rapidly the overlap decreases, consider a configuration of
two bubble walls, with $\phi \approx \mu$ between the walls, and 
$\phi \approx -\mu$ everywhere else. Except for the case when the second wall
forms very close to the first, we expect the configuration to expand at the
rate of the expansion of the universe, so that the comoving volume occupied
by $\phi \approx \mu$ remains
to a good accuracy constant, equal to some ${\cal V}$. We have confirmed this
expectation numerically for the special case of two concentric bubbles.
Let us consider the decrease in the overlap due to the displacement of
the constant mode of the field from $\phi = -\mu$ to $\phi = \mu$ in 
volume ${\cal V}$ and neglect possible displacements of all 
other modes (radiation). This assumption can only lead to an overestimate 
of the overlap. 

The constant mode has a time-dependent frequency $\omega_0(\eta)$ (with respect to 
conformal time $\eta$) given by
\be
\omega^2_0(\eta) = 2\mu^2 a^2(\eta) - 1/ \eta^2 \approx 2 \mu^2 a^2(\eta) \; ,
\ee
where $a = -1/H\eta$ is the scale factor.
Because we consider the case $\mu \gg H$, the frequency squared is
positive, so there is no amplification of the constant mode, and we can simply 
compute the overlap between the ground states corresponding to $\phi=\pm \mu$
in volume ${\cal V}$. To represent the constant mode as 
a quantized oscillator, we switch to the 
conformal field $\phi_{\rm conf} = \phi a$. The field displacement
is now $\Delta \phi_{\rm conf} = 2\mu a$. With the help of the usual formulas
for harmonic oscillators, we find the overlap of the two ground states:
\be
|\langle \phi = \mu | \phi = -\mu \rangle|^2  = 
\exp \left[ -  {\cal V} \omega_0 (\Delta \phi_{\rm conf})^2 / 2 \right] \; .
\ee
Thus, the overlap decreases to zero as an exponential of $a^3$.\footnote{
This should be contrasted with the behavior of massive gauge
theories in flat spacetime. 
There, a periodic instanton (in theories that have those) describes a transition 
between two coherent states of the field, while the subsequent real-time evolution 
corresponds to radiation of particles to infinity \cite{periodic}. 
The overlap of the final radiation state with the vacuum, computed for some cases
in ref. \cite{periodic}, remains finite, although exponentially 
small, at arbitrarily large times.}

To obtain the tunneling amplitude, we should explicitly take into account 
the possibility that a bubble can nucleate at different moment of time.
Because the ``static'' metric (\ref{met3}) is independent of $\tau$, 
there are instanton solutions obtained from the previously found
instantons by shifting $\tau$ by a (generally, complex) constant. Thus,
we consider a shift
\be
\tau \to \tau - \tau_{\rm nucl} = \tau - i t_{\rm nucl} \; ,
\ee
with an arbitrary real $t_{\rm nucl}$ as a parameter. The shifted solution has turning
points at $\tau = i t_{\rm nucl}$ and $\tau = \pi + i t_{\rm nucl}$, so that the
turning point hyperbola (\ref{hyper}) generalizes into
\be
\eta^2 - \rho^2 = \exp(-2t_{\rm nucl}) \equiv \eta_{\rm nucl}^2 \; .
\label{cosmic}
\ee
The amplitude of tunneling per a small interval of time is now
\be
\Omega d\tau_{\rm nucl} = i \Omega dt_{\rm nucl} \; ,
\ee
where $\Omega \propto \exp(-S_E)$, and $S_E$ is the action for half a period of
the periodic instanton.
The quantity $\eta_{\rm nucl}$ can be interpreted as the moment of nucleation in 
conformal time. Then, then according to (\ref{cosmic}), $t_{\rm nucl}$ is the same 
moment in time  $t = - \ln |\eta|$, which is the conventionally defined 
cosmic time (in units of $H^{-1}$).

Decoherence time can now be calculated by standard techniques. We define
$P_L(t)$ and $P_R(t)$ as the probabilities to find the field in a chosen
Hubble volume to be near $-\mu$ and $\mu$, respectively.
Following ref. \cite{CL} we write $P(t) = P_L(t) - P_R(t)$ as
\be
P(t) = \sum_{n=0}^{\infty} (-1)^n \Omega^{2n}
\int_0^t dt_{2n} \int_0^{t_{2n}} dt_{2n-1} \ldots \int_0^{t_2} dt_{1}
F(t_1,t_2,\ldots , t_{2n}) \; .
\label{series}
\ee
Here $F(t_1,t_2,\ldots , t_n)$ is the ``influence functional'' \cite{FV}, equal
to the overlap between two quantum states, obtained from a given initial state
(vacuum, in our case) in the course of two different bubble nucleation histories.
The field in the chosen Hubble volume is near $\pm \mu$ for all intervals between
the nucleation times $t_1, t_2,\ldots , t_{2n}$. However, while for 
$t_{2i} < t < t_{2i+1}$
the field is in the same well for both histories (a diagonal state),
for $t_{2i-1} < t < t_{2i}$ the wells are different (an off-diagonal state).

We have seen that if the system remains in an off-diagonal state long enough
to allow formation of two well-separated bubbles, corresponding to transitions at
$t= t_{2i-1}$ and $t = t_{2i}$, the 
overlap rapidly decreases with $t$. This means that the main contribution to
(\ref{series}) comes from pairs of histories for which the transitions at
$t_{2i-1}$ and $t_{2i}$ are ``confined'', i.e. occur back to back. For such pairs
of histories, we can approximate the influence functional as
\be
F(t_1,t_2,\ldots , t_{2n}) = \prod_{i=1}^n 
\frac{1}{M} \delta(t_{2i} - t_{2i-1}) \; ,
\label{inff}
\ee
where $M^{-1}$ is a characteristic timescale, which will be discussed shortly.
Eq. (\ref{inff}) is analogous to the ``noninteracting blip'' approximation of
ref. \cite{CL}.

Substituting (\ref{inff}) into (\ref{series}), we obtain
\be
P(t) = \sum_{n=0}^{\infty} (-1)^n \frac{\Omega^{2n}}{M^n} 
 \int_0^t dt_{2n} \int_0^{t_{2n}} dt_{2n-2} \ldots \int_0^{t_4} dt_{2} =
\exp\left( - t \Omega^2 / M \right) \; .
\label{inc}
\ee
Eq. (\ref{inc}) describes a completely incoherent (i.e. non-oscillatory)
relaxation of the probability, with decoherence time
\be
t_d = M / \Omega^2 \; .
\ee
This value sets the timescale on which, due to uncorrelated transitions
from $\phi\approx -\mu$ to $\phi\approx \mu$ and back, the probabilities to observe 
the system in either of these states become roughly equal, 
$P_L \approx P_R \approx 1/2$.

In the back-to-back events, the bubble walls form so close to each other that
they annihilate, giving rise to radiation of $\phi$ waves. (We have confirmed 
numerically that this is indeed what happens in the special case of closely 
spaced concentric bubbles.) The overlap of the quantum states, corresponding to
a history that contains such an event and a history that does not, depends
exponentially on the number of radiated particles. We therefore expect that the
main contribution to $P(t)$ will come from those back-to-back events for
which there is little or no radiation produced. This suggests that a good 
approximation for the timescale $M^{-1}$ can be obtained if we interpret each
$\Omega^2 / M$ factor in (\ref{inc}) as corresponding to a {\em complete}
periodic instanton, with the field going from one turning point to the other 
and then back. Then, $M$ can be computed from the determinant of small 
fluctuations near the periodic instanton.

Although these results were obtained for a strictly degenerate potential, we
expect them to remain practically unchanged in the nearly degenerate case 
defined by the condition (\ref{crit}). Indeed, the main distinguishing feature 
between these two cases, the existence or nonexistence of
a large critical droplet in flat space, seems to be totally irrelevant for the
calculation of decoherence time.

\section{Conclusion}
The existence of turning points for $O(4)$ symmetric instantons in de Sitter
spacetime may seem almost obvious (in view of relations (\ref{cos_zeta})). 
Nevertheless,
it is this property that distinguishes them from instantons responsible for the
$\theta$-vacuum structure in gauge theories and two-dimensional sigma-models.
For example, the $O(2)$ nonlinear sigma-model contains a single complex 
scalar field, which is not affected by conformal transformation,
and if the metric on the two-sphere is chosen in the form (\ref{met1}) (where 
now $\mu,\nu = 1,2$) the single- and multi-instanton solutions have
exactly the same form as in the flat Euclidean space. For the latter case,
they are given in ref. \cite{BP}. In particular, 
the one-instanton solution centered at the origin can be written 
as $w_1 = \irho z$,
where $z = x^1 + i x^2$, and $\irho$ is a complex parameter
(its modulus determines the instanton size). Writing 
$z = \e^{i\phi} \tan\frac{\theta}{2}$,
where $\theta$ and $\phi$ are the usual spherical angles, we can define
the ``static'' coordinates, analogous to (\ref{r}), (\ref{tau}):
\begin{eqnarray}
X & = & \cos\theta \; , \\
\tau & = & \phi \; ,
\end{eqnarray}
with $X$ running from $-1$ to 1, and $\tau$ from 0 to $2\pi$.
In these coordinates, the one-instanton solution takes the form
\be
w_1 = \irho \e^{i\tau} \left( \frac{1-X}{1+X} \right)^{1/2}  \; .
\ee
The real and imaginary parts of $w_1$ never vanish simultaneously; hence,
the solution has no turning points. Similar considerations apply to the
BPST instanton \cite{BPST} in four-dimensional massless gauge theories.

Classically, the instanton action, in either of these cases, is exactly the same
as in the flat space. We expect that the inclusion of quantum corrections will
result in a tunneling amplitude that is
somewhat reduced compared to the flat space. This is because, as in the model of
Sect. 2, the maximal instanton size is limited by the Hubble radius.
Thus, for example, for the BPST instanton, we expect the amplitude to be
suppressed by $\exp[-8\pi^2 / g^2(H)]$,  similarly to the suppression of instantons 
at finite temperature \cite{finiteT}.
Nevertheless, the resulting $\theta$-dependence
of the partition sum is nonzero and represents an effect of quantum coherence.
In contrast, instantons with turning points (periodic instantons) describe, 
as we have seen above, completely incoherent tunneling. 

The existence of turning points suggests that the instanton solutions found in
Sects. 2 and 3 can be continued to the Lorentzian signature, and the turning
points will provide initial conditions for the real-time evolution. 
To explore this real-time evolution we have made a simultaneous transformation,
eqs. (\ref{rho}), (\ref{eta}), of the temporal and radial coordinates, such that
the new coordinates are the conventional conformal coordinates of de Sitter 
spacetime. In these coordinates, the initial conditions are 
imposed on the hyperbola in the time-radius plane, see eq. 
(\ref{hyper}) and Fig. 2, and our main results were obtained by studying
this initial value problem.

These results are as follows. (i) We have found that the field that starts out
as a turning point on the hyperbola (\ref{hyper}) develops, by the time it reaches 
a section of constant conformal time $\eta=\eta_0$, into a super-Hubble
droplet of the new phase.
(ii) We have argued from causality and confirmed by numerical integration that a
super-Hubble droplet will become an expanding
bubble even in cases (such as degenerate or nearly degenerate potentials) when there
are no expanding bubbles in the flat space or such bubbles must have very large sizes.
(iii) We have shown that super-Hubble effects caused by the
expansion of de Sitter bubbles completely wash out any coherent quantum oscillations
that one might associate with a double-well potential, and we have calculated the
corresponding decoherence time.

As we have mentioned in the introduction, these results find an immediate application
in cosmological scenarios in which there are many ``worlds'' that might be connected
by quantum coherences. What we show, in a nutshell, is that a de Sitter
(or quasi-de Sitter: for example, inflationary) stage, while enhancing tunneling 
in many cases, at the same time completely destroys any traces of such coherences.
In this way, the assumption that a de Sitter stage existed in the distant past 
removes the 
problems of interpretation presented by a coherent superposition of states of the
entire universe. It also alleviates possible worries about catastrophic events
that might be thought to result if the universe today somehow picked up the ``wrong'' 
component of a superposition.

The author thanks V. Rubakov for comments on the manuscript and M. Shaposhnikov
for discussion of the results. He also thanks the Institute of Theoretical Physics,
University of Lausanne, where this work was completed, for hospitality.
This work was supported 
in part by the U.S. Department of Energy through Grant DE-FG02-91ER40681 
(Task B).


\begin{thebibliography}{99}
\bibitem{HM} S. W. Hawking and I. G. Moss, Phys. Lett. 110B, 35 (1982).
\bibitem{GW} A. H. Guth and E. J. Weinberg, Nucl. Phys. B212, 321 (1983).
\bibitem{JS} L. G. Jensen and P. J. Steinhardt, Nucl. Phys. B237, 176 (1984).
\bibitem{Starobinsky} A. A. Starobinsky, in: {\em Field Theory, Quantum Gravity,
and Strings}, eds. H. J. de Vega and N. S\'{a}nchez (Springer, 1986).
\bibitem{GL} A. S. Goncharov and A. D. Linde, Sov. J. Part. Nucl. 17, 369 (1986).
\bibitem{Deruelle} N. Deruelle, Mod. Phys. Lett. A 4, 1297 (1989).
\bibitem{RS} V. A. Rubakov and S. M. Sibiryakov, Theor. Math. Phys. 120, 1194
(1999) [gr-qc/9905093].
\bibitem{GT} S. Gratton and N. Turok, Phys. Rev. D 63, 123514 (2001).
\bibitem{CL}  S. Chakravarty and A. J. Leggett, Phys. Rev. Lett. 52, 5 (1984);
A. J. Leggett {\it et al.}, Rev. Mod. Phys. 59, 1 (1987).
\bibitem{Habib} S. Habib, Phys. Rev. D 46, 2408 (1992).
\bibitem{theta} C. G. Callan, R. F. Dashen, and D. J. Gross, Phys. Lett. 63B,
334 (1976); R. Jackiw and C. Rebbi, Phys. Rev. Lett. 37, 172 (1976).
\bibitem{etern} A. D. Linde, Phys. Lett. B 175, 395 (1986).
\bibitem{GH} G. W. Gibbons and S. W. Hawking, Phys. Rev. D 15, 2752 (1977).
\bibitem{Coleman&DeLuccia} S. Coleman and F. De Luccia, Phys. Rev. D 21, 3305 (1980).
\bibitem{FL} S. Fubini, Nuovo Cim. 34A, 521 (1976);
L. N. Lipatov, Sov. Phys. JETP 45, 216 (1977).
\bibitem{periodic} S. Khlebnikov, V. Rubakov, and P. Tinyakov, 
Nucl. Phys. B367, 334 (1991).
\bibitem{sphal} F. R. Klinkhamer and N. S. Manton, Phys. Rev. D 30, 2212 (1984).
\bibitem{tHooft} G. 't Hooft, Phys. Rev. D 14, 3432 (1976).
\bibitem{ninf_Linde} A. D. Linde, Phys. Lett. 108B, 389 (1982).
\bibitem{ninf_AS} A. Albrecht and P. J. Steinhardt, Phys. Rev. Lett. 48, 1220 (1982).
\bibitem{jumps_up} A. D. Linde, Phys. Lett. B 131, 330 (1983).
\bibitem{FV} R. P. Feynman and F. L. Vernon, Ann. Phys. (N.Y.) 24, 118 (1963).
\bibitem{BP} A. A. Belavin and A. M. Polyakov, JETP Lett. 22, 245 (1975).
\bibitem{BPST} A. A. Belavin {\em et al.}, Phys. Lett. 59B, 85 (1975).
\bibitem{finiteT} D. J. Gross, R. D. Pisarsky, and L. G. Yaffe, Rev. Mod. Phys.
53, 43 (1981).
\end{thebibliography}
\end{document}